\documentstyle[preprint,prb,aps,psfig]{revtex}
\begin{document}

%
%

\draft

\title{Energy-level statistics at the metal-insulator transition in
  anisotropic systems}

\author{Frank Milde, Rudolf A. R\"{o}mer, and Michael Schreiber}
\address{Institut f\"{u}r Physik, Technische Universit\"{a}t, D-09107
  Chemnitz, Germany}

\date{$Revision: 3.1 $; printed \today} \maketitle

\begin{abstract}
  We study the three-dimensional Anderson model of localization with
  anisotropic hopping, i.e., weakly coupled chains and weakly coupled
  planes. In our extensive numerical study we identify and
  characterize the metal-insulator transition using energy-level
  statistics. The values of the critical disorder $W_c$ are consistent
  with results of previous studies, including the transfer-matrix
  method and multifractal analysis of the wave functions.  $W_c$
  decreases from its isotropic value with a power law as a function of
  anisotropy.  Using high accuracy data for large system sizes we
  estimate the critical exponent $\nu=1.45\pm0.2$.  This is in
  agreement with its value in the isotropic case and in other models
  of the orthogonal universality class. The critical level statistics
  which is independent of the system size at the transition changes
  from its isotropic form towards the Poisson statistics with
  increasing anisotropy.
\end{abstract}


\narrowtext
\tightenlines

%
%

\section{Introduction}

It is well known that the isotropic Anderson model of
localization\cite{And58} exhibits a metal-insulator transition (MIT)
for spatial dimensions larger than two.\cite{AbrALR79,KraM93,SchG96} A
critical amount of disorder $W_c$ is necessary to localize all the
eigenstates. The asymptotic envelopes of the localized states for
$W>W_c$ decay exponentially in space due to the destructive
interference of the disorder-backscattered wave functions with
themselves. An electron confined in such a state cannot contribute to
charge transfer at temperature $T=0$. For $W<W_c$ there exist states
that are extended through the whole sample allowing charge transfer
through the system at $T=0$.  In spatial dimensions up to two an
arbitrarily small amount of disorder localizes all states and there is
no MIT at finite $W$ for non-interacting
systems.\cite{AbrALR79,KraM93}

In the present study we consider the problem of Anderson localization
in three dimensional (3D) disordered systems with anisotropic hopping.
One might expect that increasing the hopping anisotropy, namely
reducing the hopping probability in one ore two directions, is
effectively equivalent to changing the dimensionality of the system
continuously from 3D to 2D or 1D similarly to the case of (bi)-fractal
lattices.\cite{SchG96} However, this is not the case: The topology of
the lattice is not changed as long as the hopping is non-zero and the
dimensionality is still three. Only if the coupling is reduced to zero
the dimension jumps from 3D to 2D or 1D.  Previous studies using the
transfer-matrix method (TMM)\cite{LiSEG89,ZamLES96a,PanE94} and
multifractal analysis (MFA)\cite{MilRS97} showed that an MIT exists
even for very strong anisotropy.  The values of the critical disorder
$W_c$ were found to decrease by a power law in the anisotropy,
reaching zero only for the limiting 1D or 2D cases. In the present
work, we focus our attention on the critical properties of this MIT.
As an independent check to previous results, we employ energy level
statistics (ELS)\cite{EvaE92,ShkSSL93,HofS93,HofS94b} together with
the finite-size scaling (FSS) approach. ELS has been previously
applied with much success at the MIT of the isotropic model and it was
shown that a size-independent statistics exists at the
MIT.\cite{ShkSSL93,HofS94b,ZhaK97} This {\em critical} statistics is
intermediate between the two limiting cases of Poisson statistics for
the localized states and the statistics of the Gaussian orthogonal
ensemble (GOE) which describes the spectrum of extended
states.\cite{Meh90} We check whether the critical ELS is influenced by
the anisotropy.

A major part of our study is dedicated to the determination of the
critical exponent $\nu$ of this second-order phase transition. In
general it is assumed that $\nu$ depends only on general symmetries,
described by the universality class, but not on microscopic details of
the sample.  Thus, anisotropic hopping might shift $W_c$ but should
not change $\nu$. In order to verify this assumption it is necessary
to determine $\nu$ with high accuracy for various anisotropies.  This
is a computationally demanding task and we emphasize that even in
numerical studies of the isotropic Anderson model the correct value of
$\nu$ is still not entirely clear. Recent highly accurate TMM studies
report $\nu=1.54\pm0.08$\cite{Kin94} and
$\nu=1.58\pm0.06$\cite{SleO99a}, but from ELS
$\nu=1.35\pm0.15$\cite{Hof98} and $\nu=1.4\pm0.15$\cite{ZhaK97} was
found. Also from TMM $\nu=1.3\pm0.1$ for the isotropic system and
$\nu=1.3\pm0.1$ and $\nu=1.3\pm0.3$ for two different data sets for
the anisotropic system were determined.\cite{ZamLES96a} As we will
show, an accurate estimation of $\nu$ requires the computation of
large system sizes before FSS can be reliably employed.  Using high
precision data and taking into account non-linear corrections to
scaling\cite{SleO99a} we then find $\nu= 1.45\pm 0.2$.

The paper is organized as follows. In Sec.~\ref{sec-model} we
introduce our notation. We recall the use of ELS for the
characterization of the MIT in Sec.~\ref{sec-ELS}. Using the fact that
the statistical properties at the transition do not depend on the
system size, we corroborate the existence of the MIT and determine the
anisotropy dependence of the critical disorder and of the critical
statistics.  In particular, we show that the critical ELS changes
continuously with increasing anisotropy from its functional form at
the isotropic MIT towards Poisson statistics. In
Sec.~\ref{sec-Exponent} we describe the concept of FSS and the
numerical methods to determine the critical properties such as $\nu$.
Then we demonstrate that the scaling concept applies to the integrated
$\Delta_3$ statistics from ELS\cite{Hof98} and we estimate $\nu$ from
highly accurate ELS data obtained for very large system sizes. In
Sec.~\ref{sec-Cels} we also check whether our results are compatible
with highly accurate TMM data. Finally, we summarize our results in
Sec.~\ref{sec-SUM}.

%
%

\section{The anisotropic Anderson model of localization}
\label{sec-model}

The Anderson model is a standard model for the description of
disordered systems with Hamiltonian given as\cite{And58}
\begin{equation}
  \label{Hand}
  H = \sum_{i} \epsilon_{i} | i \rangle\langle i | + \sum_{i \ne j}
  t_{ij} | i \rangle\langle j | \quad .
\end{equation}
The states $| i \rangle$ are orthonormal and correspond to particles
located at sites $i=(x,y,z)$ of a regular cubic lattice with size
$N^3$. We use periodic boundary conditions $(x+N,y,z) = (x,y+N,z) =
(x,y,z+N) = (x,y,z)$. The potential site energies $\epsilon_{i}$ are
uniformly distributed in the interval $[-W/2,W/2]$ with $W$ defining
the disorder strength, i.e., the amplitude of the fluctuations of the
potential energy. The transfer integrals $t_{ij}$ are restricted to
nearest neighbors and depend only on the three spatial directions, so
$t_{ij}$ can either be $t_x$, $t_y$ or $t_z$.  We study two
possibilities of anisotropic transport: (i) {\em weakly coupled
  planes} with
\begin{equation}
  \label{eq:gammaP}
  t_x=t_y=1$, $t_z=1-\gamma
\end{equation}
and (ii) {\em weakly coupled chains} with
\begin{equation}
  \label{eq:gammaC}
  t_x=t_y=1-\gamma$, $t_z=1 \quad .
\end{equation}
This defines the strength of the hopping anisotropy $\gamma\in [0,1]$.
For $\gamma=0$ we recover the isotropic case, $\gamma=1$ corresponds
to $N$ independent planes or $N^2$ independent chains.

%
%

\section{Energy level statistics}
\label{sec-ELS}

\subsection{ELS and MIT}

The statistical properties of the energy spectra reflect the character
of the eigenstates and have been proven to be a powerful tool for
characterizing the
MIT.\cite{EvaE92,ShkSSL93,HofS93,HofS94b,ZhaK97,Hof98} On the
insulating side of the MIT, one finds that localized states that are
close in energy are usually well separated in space whereas states
that are localized in vicinal regions in space have well separated
eigenvalues. Consequently, the eigenvalues on the insulating side are
uncorrelated, there is no level repulsion and the probability of
eigenvalues to be close together is high. This is called level
clustering and is described by the Poisson statistics.\cite{Meh90} On
the other hand, extended states occupy the same regions in space and
their eigenvalues become correlated. This results in level repulsion
such that the spectral properties are given by the GOE
statistics.\cite{Meh90}

In an infinitely large disordered system, the MIT corresponds to a
sharp transition from GOE statistics at the metallic side to Poisson
statistics at the insulating side via some intermediate critical
statistics only exactly at the critical point.\cite{ShkSSL93,HofS94b}
In a finite system, this abrupt change is smeared out, because the
divergence of the characteristic lengths --- such as the localization
length --- of the wave functions at the phase transition is cut off at
the system size.  If for a given $W$ the localization length in the
infinite system is much larger than the system size under
consideration, the states appear to be extended in the finite system.
Their eigenvalues become correlated and the ELS is changed from
Poisson towards GOE statistics.  To obtain a reliable characterization
of the MIT one should therefore investigate the system-size dependence
of the spectral properties: With increasing system size there is a
trend towards the limiting cases of GOE and Poisson statistics for the
extended and localized regions, respectively.  Directly at the
critical disorder there are no characteristic length scales, the wave
functions are scale-invariant multifractal
entities\cite{MilRS97,SchG91} and the statistical properties of the
spectrum are independent of the system size.\cite{ShkSSL93,HofS94b}

\subsection{The numerical approach}
\label{sec-ELS-numerical}

In ELS a system is characterized by the local fluctuations of the
energy spectrum around its average density of states (DOS) $\bar\rho
(E)$.\cite{Meh90} Usually, $\bar\rho (E)$ is not constant for a given
sample and has different width or even different shape for different
samples. We therefore apply a so-called unfolding
procedure\cite{HofS93} to map the set of eigenvalues $\{E_i\}$ to a
new set $\{\varepsilon_i\}$ with constant average density equal to
unity as required for the application of random matrix
theory.\cite{Meh90} We then characterize the unfolded spectrum
$\{\varepsilon_i\}$ by the distribution $P(s)$, which measures the
level repulsion in terms of the nearest-neighbor level-spacing $s$,
by the cumulative level-spacing distribution $I(s)=\int_s^\infty P(s')
ds'$, and by the $\Delta_3$ statistics, which measures the deviation
from a sequence of $L$ uniformly spaced levels,\cite{HofS93,Meh90}
\begin{equation}
  \label{eq:Delta3}
  \Delta_3(L)= \left\langle \frac{1}{L} \min_{A,B}
    \int_\varepsilon^{\varepsilon+L}
    [D(\varepsilon')-A\varepsilon'-B]^2 d\varepsilon'
  \right\rangle_\varepsilon \quad .
\end{equation}
Here, $D(\varepsilon)$ is the integrated DOS and $\langle
\rangle_\varepsilon$ corresponds to averaging over the spectrum.

For the eigenvalue computation the Lanczos algorithm in the
Cullum-Willoughby implementation\cite{CulW85a,CulW85b} is applied
which is very effective for our sparse matrices.\cite{ElsMMR99} We use
system sizes up to $N=50$ for which a 400 MHz Pentium II machine needs
about five days for the diagonalization of a single system.  The
character of the eigenstates has been shown not to change in a large
energy interval around $E=0$.\cite{BulSK87} For the computation of the
spectral properties we therefore use an interval centered at $E=0$
containing 50\% of the eigenvalues.\cite{HofS93,ZhaK97} Furthermore we
average over a number of configurations of the potential site energies
such that at least $10^5$, but typically $2\times 10^5$ to $4 \times
10^5$, eigenvalues contribute to $P(s)$, $I(s)$ or $\Delta_3(L)$ for
every $N$, $W$, and $\gamma$. Altogether we investigated about 750
such parameter combinations.


\subsection{Dependence of $W_c$ on anisotropy}
\label{sec-ELS-wc}

\begin{figure}[p]
  \centerline{\psfig{figure=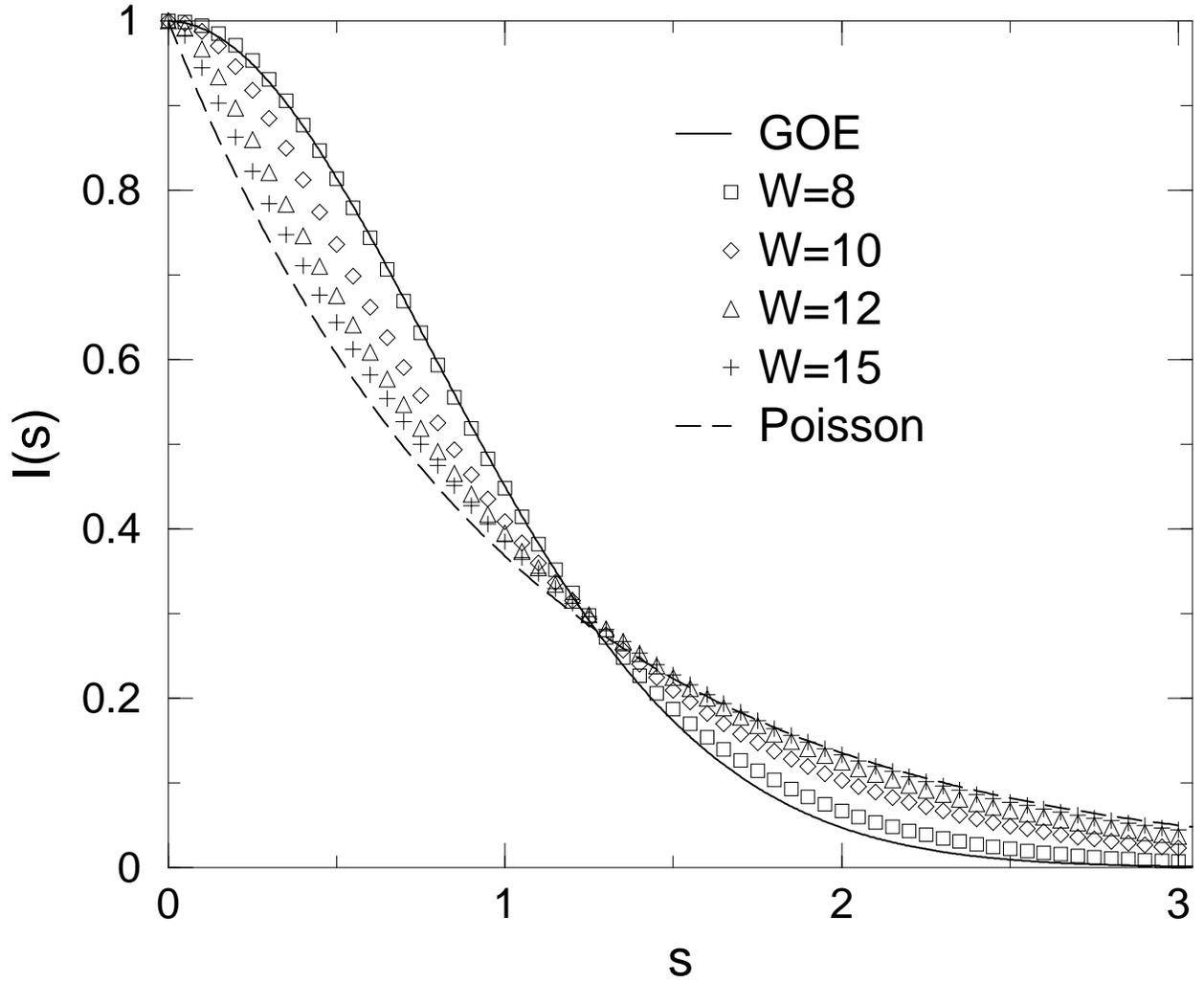}}
  \vspace{1cm}
  \caption{\label{fig:Is}
    Cumulative spacing distribution $I(s)$ for coupled chains with
    $\gamma=0.6$ and $N=24$ and various disorders. Solid and dashed
    lines indicate GOE and Poisson behavior, respectively.}
\end{figure}

\begin{figure}[p]
  \centerline{\psfig{figure=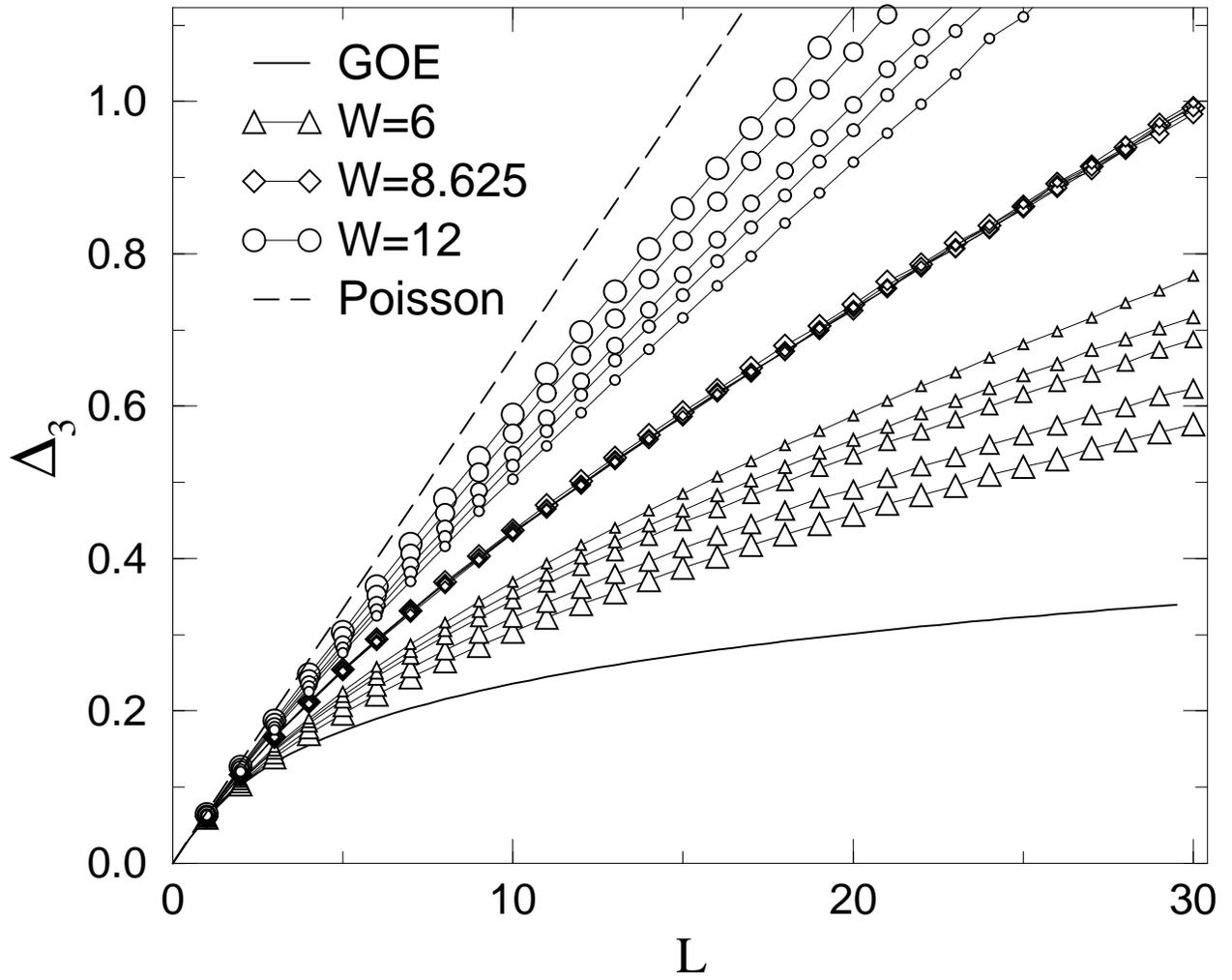}}
  \vspace{1cm}
  \caption{\label{fig:d3}
    $\Delta_3$ statistics for coupled planes with $\gamma=0.9$ and
    system sizes $N=13, 17, 21, 30, 40$, indicated by increasing
    symbol size.}
\end{figure}

\begin{figure}[p]
  \centerline{\psfig{figure=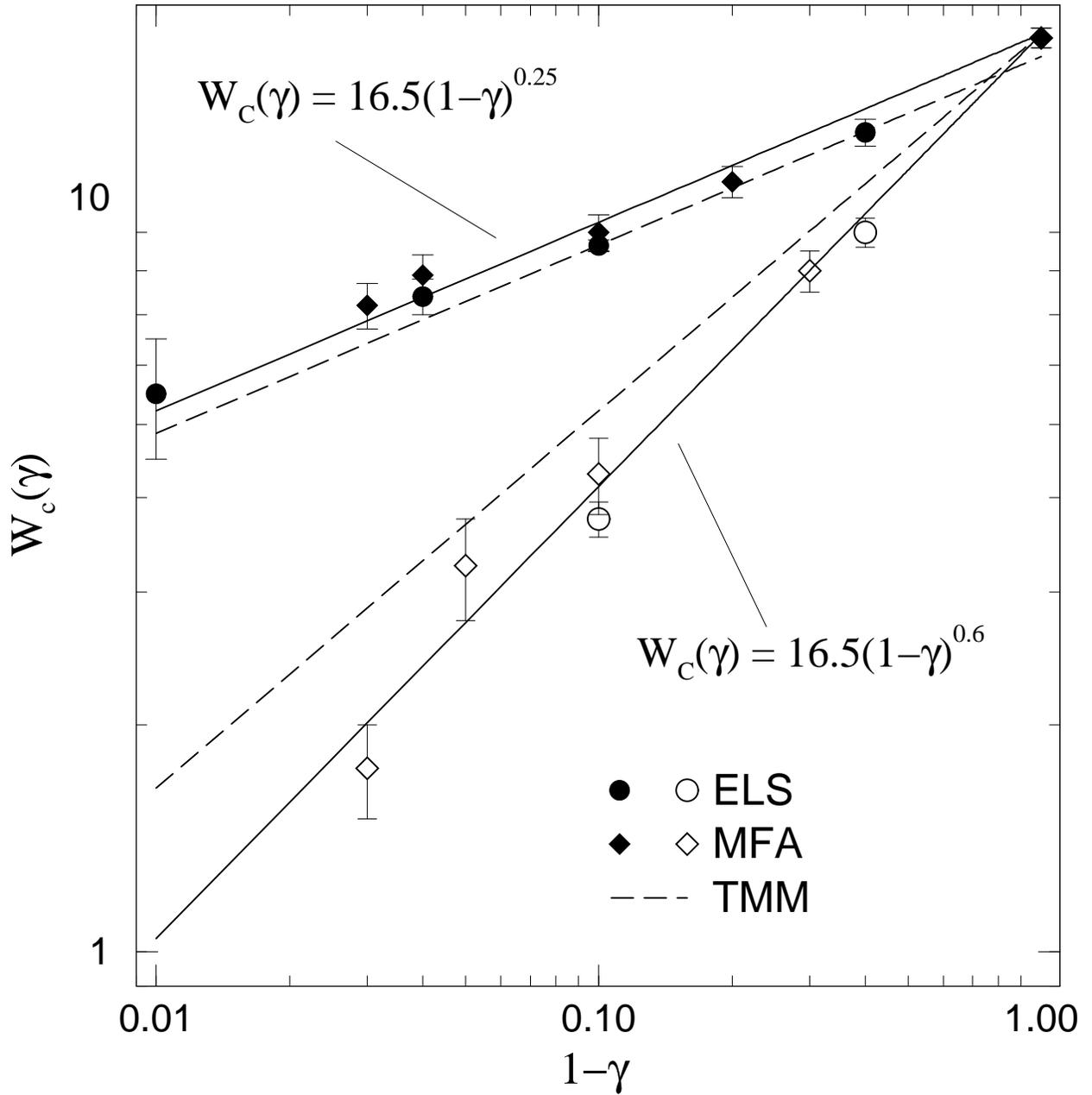}}
  \vspace{1cm}
  \caption{\label{fig:wc}
    Anisotropy dependence of $W_c$ for coupled planes (filled symbols)
    and chains (open symbols) as computed by ELS and
    previously\protect\cite{MilRS97} by MFA.  We also added a fit to
    TMM data of Ref.~\protect\onlinecite{ZamLES96a} (dashed lines).}
\end{figure}

As expected, we find a crossover from GOE statistics to Poisson
statistics with increasing $W$ for all values of $\gamma$ considered
and both, coupled planes and chains. As an example we show $I(s)$ for
weakly coupled chains in Fig.\ \ref{fig:Is}.  This crossover is a
first hint for the existence of an MIT.  In order to check this
further we investigate the system-size dependence of the $\Delta_3$
statistics.  In Fig.\ \ref{fig:d3} we show $\Delta_3(L)$ for weakly
coupled planes for various system sizes.  There is a clear trend
towards GOE and Poisson statistics for $W=6$ and $W=12$, respectively.
But there is hardly any system-size dependence visible for $W=8.625$.
As described above, this indicates an MIT with a critical disorder in
the vicinity of $W= 8.625$. For a more accurate determination of
$W_c$, we consider the integral $\alpha_N(W)=\int_0^{30}
\Delta_3(L,W,N) dL$ as a function of $W$ for several system sizes
$N$.\cite{HofS94b} As can be seen from Fig.\ \ref{fig:d3}, the value
of $\alpha_N$ monotonically increases as the ELS changes from GOE to
Poisson statistics. In the localized region $\alpha_N$ will therefore
increase with $N$, whereas it will decrease with $N$ for extended
states. One can then determine $W_c$ from plots of $\alpha_N(W)$ for
different $N$ as shown, e.g., in Fig.\ \ref{fig:alpha_W} for
$\gamma=0.9$.  All curves cross in one point, at which the size
effects change sign. This indicates the transition which occurs at
$W_c=8.6\pm0.2$ in this case.  Our results for other values of
$\gamma$ are compiled in Fig.\ \ref{fig:wc}.  We find that with
increasing anisotropy the critical disorder decreases according to a
power law $W_c=16.5 (1-\gamma)^\beta$ with $\beta\approx 0.25$ for
coupled planes and $\beta\approx 0.6$ for coupled chains. As shown in
Fig.\ \ref{fig:wc} this is an appropriate description for our ELS data
and agrees well with the results of our previous MFA.\cite{MilRS97}
For coupled planes this result is also consistent with a previous TMM
study and a perturbative analysis employing the coherent potential
approximation (CPA)\cite{ZamLES96a} but in the case of coupled chains
$\beta\approx 0.6$ appears more appropriate than the result
$\beta=0.5$ of Ref.\ \onlinecite{ZamLES96a}.

\subsection{Dependence of $P_c(s)$ on anisotropy}
\label{sec-ELS-pc}

\begin{figure}[p]
  \centerline{\psfig{figure=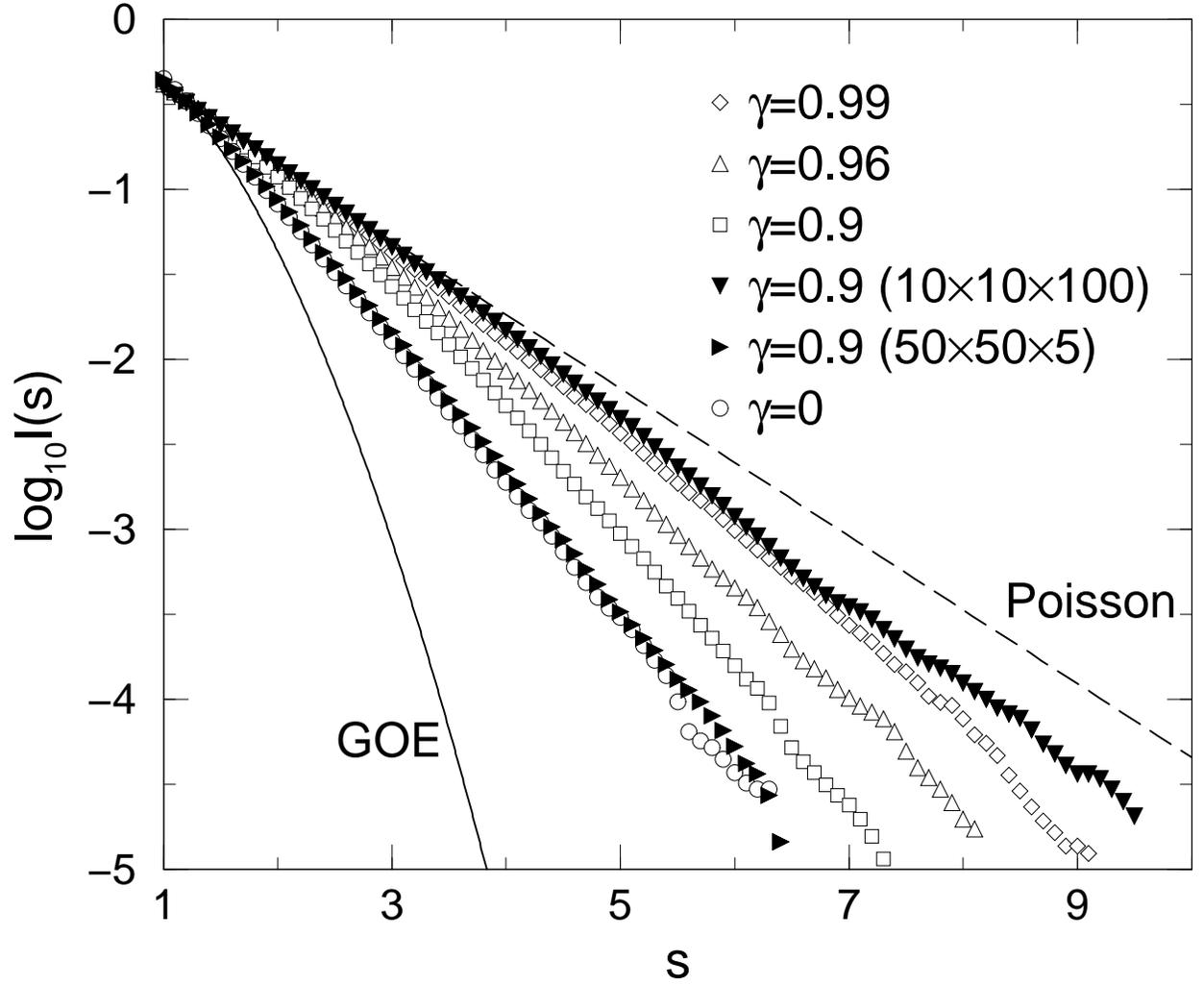}}
  \vspace{1cm}
  \caption{\label{fig:IS_KRIT}
    Large-$s$ behavior of the cumulative level spacing-distribution
    $I_c(s)$ of coupled planes at the MIT for various values of
    anisotropy $\gamma$. The filled symbols correspond to the
    non-cubic samples as discussed in Sec.\ \protect\ref{sec-ELS-pc}.}
\end{figure}

Let us now turn our attention to the question whether the form of the
size-independent statistics at the MIT $P_c(s)$ depends on anisotropy.
It seems to be settled --- at least for the isotropic case --- that
the small-$s$ behavior of $P_c(s)$ is equal to that of the metallic
phase with $P_c(s)\propto s$ as usual for the orthogonal ensemble,
$s^2$ for the unitary, and $s^4$ for the symplectic
ensembles.\cite{Hof96,BatSZK96} Furthermore, it was shown that the
large-$s$ behavior of $P_c(s)$ and $I_c(s)$ can be described by an
exponential decay $P_c(s) \propto I_c(s)\propto e^{-A_c s}$ with
$A_c\approx 1.9$ for all three universality
classes.\cite{ZhaK97,Hof98,Hof96,BatSZK96} All these studies were
performed for 3D using periodic boundary conditions and cubic samples.
On the other hand, $P_c(s)$ has been shown to depend on the sample
shape\cite{PotS98} and on the applied boundary conditions as
well.\cite{BraMP98,SchP98} A trend towards Poisson behavior was found
when the cube was deformed by increasing or decreasing the length in
one direction\cite{PotS98} or when periodic boundary conditions were
changed to Dirichlet in one, two, or three
directions.\cite{BraMP98,SchP98} Thus $A_c$ decreases from $1.9$
towards $1$. Furthermore, $A_c$ depends on the dimensionality since in
the orthogonal 4D case the critical $P_c(s)$ was found to be closer to
Poisson statistics than in 3D.\cite{ZhaK98} From the 4D results one
might expect the opposite effect for our coupled planes and chains.
This is not the case and we also find a trend towards Poisson
statistics for increasing $\gamma$ as can be seen, e.g., in Fig.\ 
\ref{fig:IS_KRIT} for coupled planes. This finding is consistent with
the MFA results\cite{MilRS97} where the singularity spectra at the
transition were found to tend towards the localized behavior with
increasing anisotropy.

While investigating the dependence on the sample shape we observe
another interesting behavior. In the isotropic case, when deforming
the cubic sample to a cuboid, the statistical properties of the
spectra always tend towards Poisson statistics, irrespective of
whether the sample becomes a long quasi-1D bar or a flat quasi-2D
sample.\cite{PotS98} Here we compute $I(s)$ for coupled planes with
$\gamma=0.9$ at $W=W_c=8.625$ for two cases: (i) bar-shaped samples of
size $10\times 10\times 100$ extending in the direction with reduced
hopping and (ii) flat samples of size $50\times 50\times 5$ with large
weakly coupled planes. We insert the results into Fig.\ 
\ref{fig:IS_KRIT}.  Surprisingly we find an opposite trend for the two
cases: (i) for the bars $I(s)$ is close to Poisson statistics, very
similar to $I_c(s)$ for $\gamma=0.99$; (ii) for the flat samples
$I(s)$ is close to GOE statistics and the isotropic $I_c(s)$ is nearly
recovered.  This result is probably due to the fact, that the system
sizes in case (ii) are proportional to the localization lengths, which
depend on the direction.\cite{ZamLES96a} The extension of the wave
function measured in units of its characteristic lengths is then equal
in all directions.  We remark that a similar observation in 2D
anisotropic samples exists: the isotropic scaling function is
recovered, if the dimensions of the system are proportional to the
localization lengths.\cite{LiKES97} We expect that a further increase
of the aspect ratio, i.e., reduction of the system towards 2D, will
drive $I(s)$ again towards Poisson statistics. But this needs further
study.

We remark, that the increasing fluctuations for $s>6$ in the $I(s)$
curves in Fig.\ \ref{fig:IS_KRIT} are due to the fact, that there are
only very few such large spacings. Consequently, there is no good
statistics. In tests with large Poisson sequences of up to $10^6$
eigenvalues, we find similar small but increasing deviations from the
theoretical result for $s>6$.

%
%
\section{One-parameter scaling at the MIT}
\label{sec-Exponent}

The MIT in the Anderson model of localization is expected to be a
second-order phase transition,\cite{AbrALR79,BelK94} which is
characterized by a divergence in an appropriate correlation length
\begin{equation}
  \label{eq:CorrLength}
  \xi_\infty(W)=C|W-W_c|^{-\nu}
\end{equation}
with critical exponent $\nu$, where $C$ is a constant.\cite{KraM93}
Here $\xi_\infty(W)$ is the correlation length of the {\em infinite}
system, but in practice only finite, and still relatively small,
systems are numerically accessibly. In order to construct a
thermodynamic limit, scaling laws $X(W,bN)=F(X(W,N),b)$ are applied to
the finite-size data $X(W,N)$. Here $X$ denotes a dimensionless system
property to be specified later and $b$ is an arbitrary scale
factor.\cite{Tho74} The scaling law has solutions of the form
\begin{equation}
  \label{eq:ScalFunc}
  X=f(N/\xi_\infty) 
\end{equation}
which implies that the system size $N$ can be scaled by
$\xi_\infty(W)$ such that all $X(W,N)$ collapse onto a single scaling
function $f$.  For a system with an MIT, this scaling function
consists of two branches corresponding to the localized and the
extended phase. In numerical experiments the reduced localization
length $\Lambda_N(W)$ obtained by the TMM is often used as quantity
$X$.\cite{KraM93,ZamLES96a,PicS81a,MacK81} Scaling has also been shown
for quantities derived from ELS, particularly for $\alpha_N(W)$
defined in Sec.\ \ref{sec-ELS-wc}.\cite{ShkSSL93,HofS94b}

For the estimation of the critical exponent $\nu$, one can numerically
perform the FSS procedure\cite{MacK83} in order to determine the
scaling function $f(N/\xi_\infty)$ by minimizing the deviations of the
data from the common scaling curve. After construction of the scaling
curve, one can then fit the obtained scaling parameters
$\xi_\infty(W)$ according to Eq.\ (\ref{eq:CorrLength}). However, the
divergence of $\xi_\infty(W)$ at $W_c$ is rounded because of numerical
inaccuracies.\cite{MacK83} On the other hand, Eq.\ 
(\ref{eq:CorrLength}) is not expected to be valid far away from $W_c$
and it is a priori very difficult to determine an appropriate disorder
range for the fit.

There are several possibilities of determining $\nu$ directly from
$X(W,N)$, thereby avoiding the numerical inaccuracies introduced by
the scaling procedure.  Linearizing Eq.\ (\ref{eq:ScalFunc}) at the
transition and using Eq.\ (\ref{eq:CorrLength}) one finds that close
to $W_c$ the quantity $X$ behaves as\cite{KraM93}
\begin{equation}
  \label{eq:X_W_Dep}
  X(W,N)=X(W_c,N)+\tilde{C}(W-W_c)N^{1/\nu} \quad .
\end{equation}
Fitting the linear range of $X(W)$ for various fixed values of $N$ now
allows us to determine $\nu$. One can also find a similar expression
where $X$ is replaced by $\ln{X}$.\cite{Kin94} Although both
expressions are equivalent close to the MIT, they can give different
results due to finite precision numerics.  Another complication
appears due to the presence of a systematic shift of the crossing
point of the $X(W)$ curves visible, e.g., in highly accurate TMM
data.\cite{Kin94,SleO99a} Such a shift occurs also in our ELS data,
but it is less prominent than in the mentioned TMM studies and can
barely be seen in the inset of Fig.\ \ref{fig:alpha_W}. The shift is
not described by Eq.\ (\ref{eq:X_W_Dep}). The first attempt to
overcome this problem was adding a correction term $B$ to Eq.\ 
(\ref{eq:X_W_Dep}) which depends on $N$ but not on $W$.\cite{Kin94}
This correction allows us to determine $\nu$ without an assumption
about the nature of the shift.  However, the value of $W_c$ is not
accessible in this procedure since a change in $W_c$ can be
compensated by changing the correction $B(N)$ accordingly.

Alternatively one can assume that the small deviations from
one-parameter scaling are caused by an irrelevant scaling variable,
i.e., the presence of additional terms in (\ref{eq:X_W_Dep}) with
system-size dependence $N^{y}$, with $y<0$, which vanish for large
system sizes.  This approach takes into account that the shift is not
random but rather appears to be systematic.  We use such a method
introduced recently.\cite{SleO99a} A family of fit functions is
constructed, which includes two kinds of corrections to scaling: (i)
an irrelevant scaling variable and (ii) nonlinearities of the disorder
dependence of the scaling variables.  Starting point is the
renormalization group equation (or scaling function) for $X$
\begin{equation}
  \label{eq:SlevenRenorm}
  X=\tilde{f}(\chi_{\rm r} N^{1/\nu}, \chi_{\rm i} N^{y}) \quad .
\end{equation}
$\chi_{\rm r}$ and $\chi_{\rm i}$ are the relevant and irrelevant
scaling variables with corresponding critical and irrelevant exponents
$\nu$ and $y$, respectively. $\tilde{f}$ is then Taylor expanded up to
order $n_{\rm i}$ in terms of the second argument
\begin{equation}
  \label{eq:SlevenRenorm2}
  X=\sum_{n=0}^{n_{\rm i}} \chi_{\rm i}^n N^{n
    y}\tilde{f}_n(\chi_{\rm r} N^{1/\nu}) \quad ,
\end{equation}
and each $\tilde{f}_n$ is Taylor expanded up to order $n_{\rm
  r}$:
\begin{equation}
  \tilde{f}_n=\sum_{i=0}^{n_{\rm r}} a_{ni} \chi_{\rm r}^i N^{i/\nu}
  \quad .
\end{equation}
Finally, nonlinearities are taken into account by expanding
$\chi_{\rm r}$ and $\chi_{\rm i}$ in terms of $w=(W_c-W)/W_c$ up to
order $m_{\rm r}$ and $m_{\rm i}$, respectively,
\begin{equation}
  \label{eq:SlevenScalVar}
  \chi_{\rm r}(w)=\sum_{n=1}^{m_{\rm r}} b_n w^n , \quad
  \chi_{\rm i}(w)=\sum_{n=0}^{m_{\rm i}} c_n w^n \quad ,
\end{equation}
with $b_1=c_0=1$.  Choosing the orders $n_{\rm i}, n_{\rm r}, m_{\rm
  r}, m_{\rm i}$ up to which the expansions are carried out, one can
adjust the fit function to the data set. If there is a single crossing
point, an irrelevant scaling variable is not necessary and $n_{\rm i}$
is set to zero. In order to recover the linear behavior
(\ref{eq:X_W_Dep}), one chooses $n_{\rm r}=m_{\rm r}=1$ and $n_{\rm
  i}=m_{\rm i}=0$. However, we emphasize that the linear region around
$W_c$ is very small and a simple scaling with Eq.\ (\ref{eq:X_W_Dep})
will not give accurate results. For larger $W$ ranges, $2$nd or $3$rd
order terms are necessary for the fit.  Setting $n_{\rm r}$ or $m_{\rm
  r}$ to two or three yields appropriate fit functions.  The total
number of fit parameters including $W_c$, $\nu$, and $y$ is
$N_p=(n_{\rm i}+1)(n_{\rm r}+1)+m_{\rm r}+m_{\rm i}+2$ and should of
course be kept as small as possible.

There is an alternative method developed for ELS which should give very
accurate results with a relatively small number of data
points.\cite{Hof98} The data $X(W)$ for constant $N$ are fitted with
third-order polynomials and these functions are used to generate a
large number of new data points for which the FSS
procedure\cite{MacK83} is employed. This saves a lot of computer time
but does not solve the problems of the FSS procedure discussed above.
Furthermore, this smoothing pretends a higher quality of the data than
actually achieved.

%
%
\section{Computation of the critical properties at the MIT}
\label{sec-Cels}

We have decided to determine the critical exponent $\nu$ for coupled
planes with strong anisotropy $\gamma=0.9$ with highest accuracy.  We
therefore doubled the number of samples for this value of $\gamma$
compared to the other cases and used large system sizes, namely
$N=13,17,21,24,30,40$, and $50$. The disorder range is $W\in [6,12]$
accept for $N=50$ where we reduced it to $W\in[8,9.25]$.
These data are shown in Fig.\ \ref{fig:alpha_W}. Here, $5\times10^5$
to $7 \times10^5$ eigenvalues contribute to each $\alpha_N(W)$ and the
statistical error from the average over the $\alpha_N$ values from
each sample is between 0.2\% and 0.4\%. The number of samples ranges
from $699$ for $N=13$ to $10$ for $N=50$.

\subsection{The non-linear fit}

For the non-linear fit to the data, we use the Levenberg-Marquardt
method\cite{PreFTV92} (LMM) as implemented, e.g., in the {\em
  Mathematica} function {\tt NonlinearRegress}.  The LMM minimizes the
$\chi^2$ statistics, measuring the deviation between model and data
under consideration of the error of the data points. In order to judge
the quality of the fit, we use the goodness-of-fit parameter
$Q$,\cite{PreFTV92} which incorporates besides the value of $\chi^2$
also the number of data points and fit parameters. For reliable fits
$Q$ should fall into the range $0.01<Q<1$.\cite{PreFTV92} The output
of the LMM routine contains also confidence intervals for the
estimated fit parameters.  We check these independently by computing a
number of new data sets by randomly varying each data point
$\alpha_N(W)$ within its error bar. The LMM fit procedure is then
applied to these new sets and the variance of the resulting $W_c$ and
$\nu$ values is compared to the confidence intervals of the original
fit. We find that they do not differ significantly. More importantly,
we also check whether the fit parameters are compatible when using
different expansions of the fit function as outlined in the last
section. As we will show below, it turns out that then the fitted
values of $W_c$ and $\nu$ differ from each other by more than the
confidence intervals obtained from the individual LMM fits. This is
important when coalescing all of them into a final result.

\subsection{Results}
\label{sec-Cels-res}

\begin{figure}[p]
  \centerline{\psfig{figure=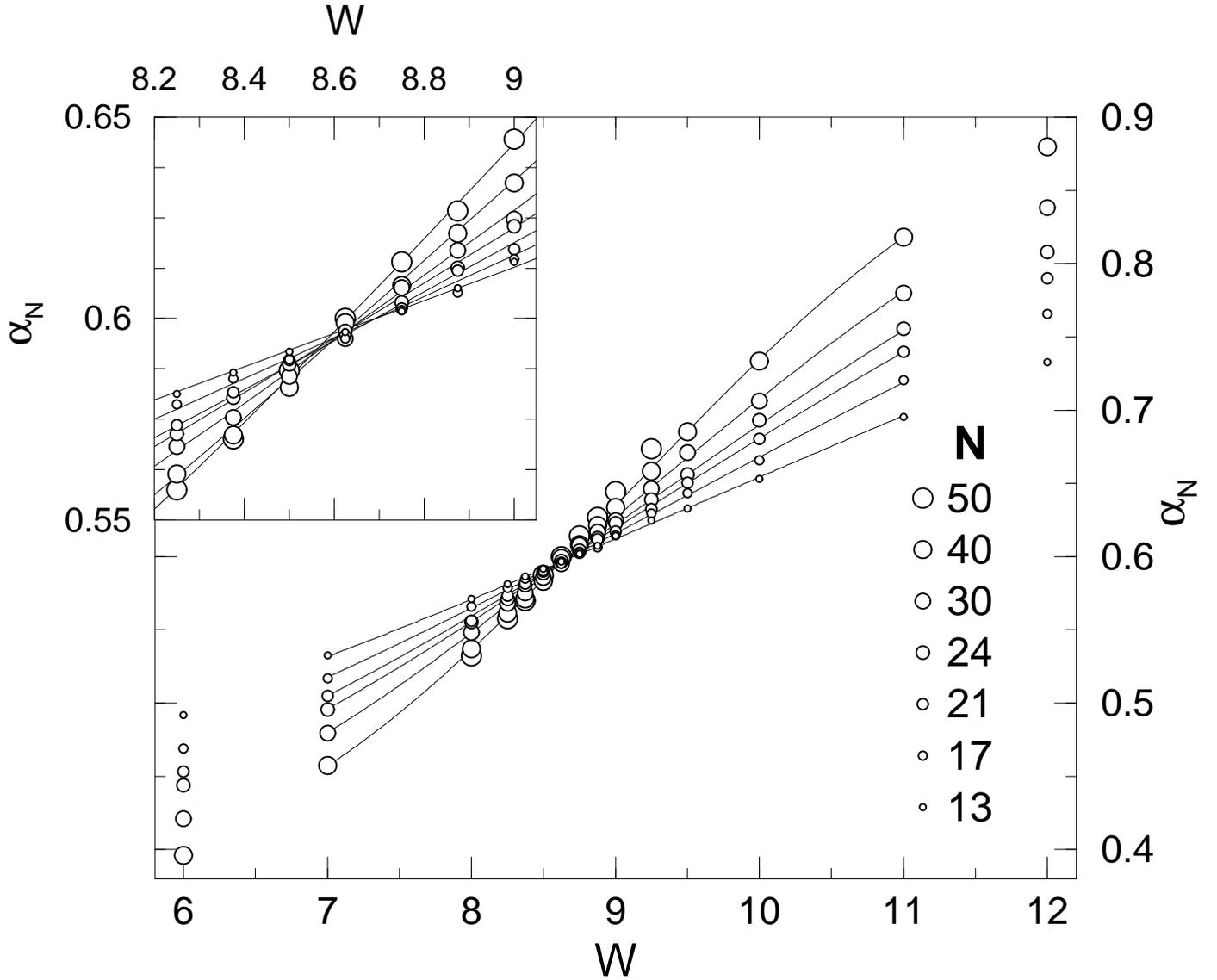}}
  \vspace{1cm}
  \caption{\label{fig:alpha_W}
    The integrated $\Delta_3$ statistics $\alpha_N(W)$ for coupled
    planes with $\gamma=0.9$ as a function of disorder $W$ for various
    system sizes $N$. The lines correspond to fit A of Table
    \protect\ref{tab:fitdata_ELS}. In the inset we show enlarged the
    region around the crossing point. Here, the lines are cubic fits
    to the data.}
\end{figure}

\begin{figure}[p]
  \centerline{\psfig{figure=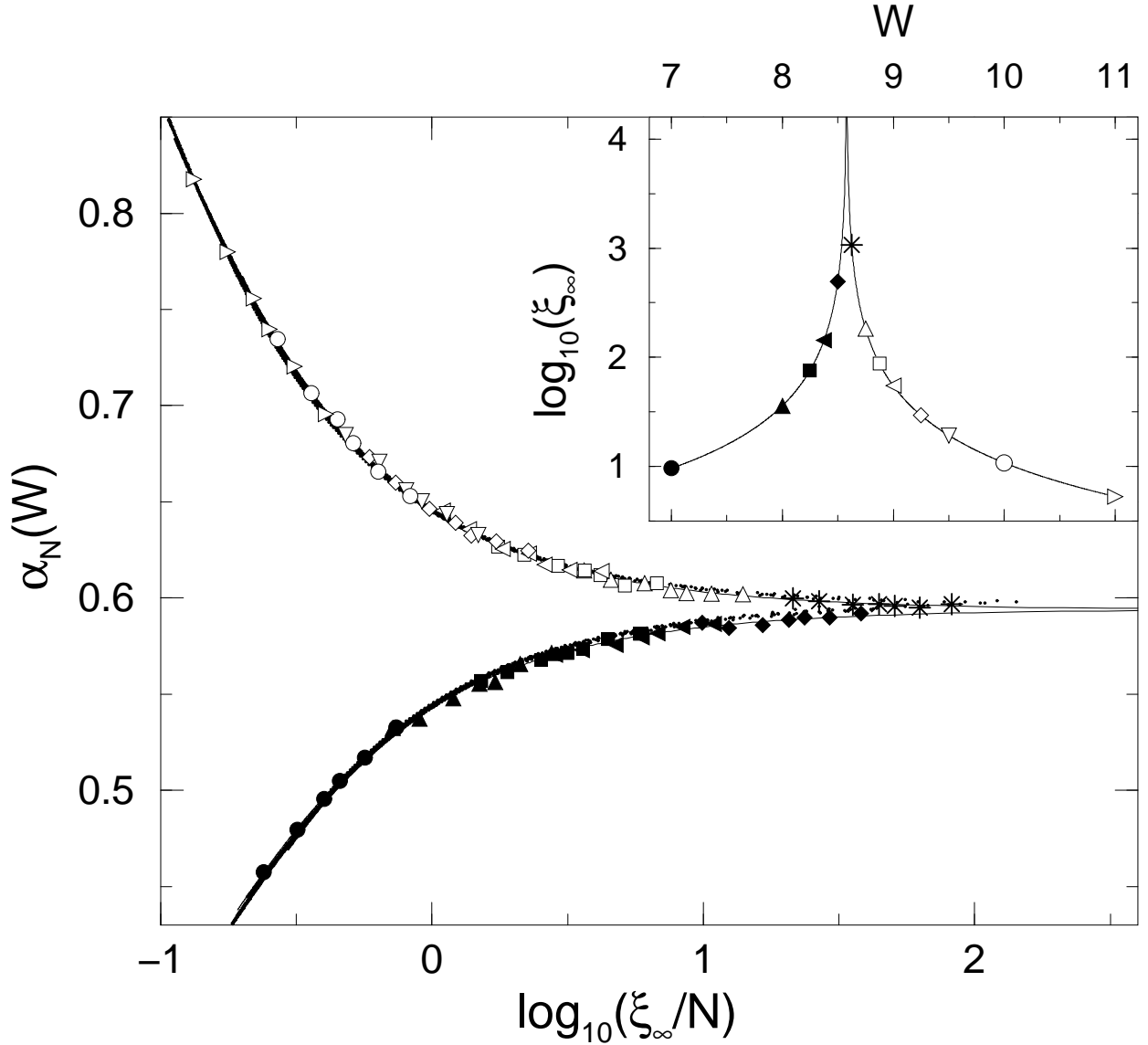}}
  \vspace{1cm}
  \caption{\label{fig:scalf_alpha}
    Scaling function $\alpha_N=f(\xi_\infty/N)$ and scaling parameter
    $\xi_\infty$ (inset) from fit f in Table
    \protect\ref{tab:fitdata_ELS} with localized (open symbols) and
    extended branch (filled symbols). Different symbols distinguish
    the data for different $W$. The lines are the functional forms as
    given by the fit. We added a scaling function (dots) obtained by a
    different method.\protect\cite{Hof98} The large number of dots
    partly results in a broad line.}
\end{figure}

\begin{figure}[p]
  \centerline{\psfig{figure=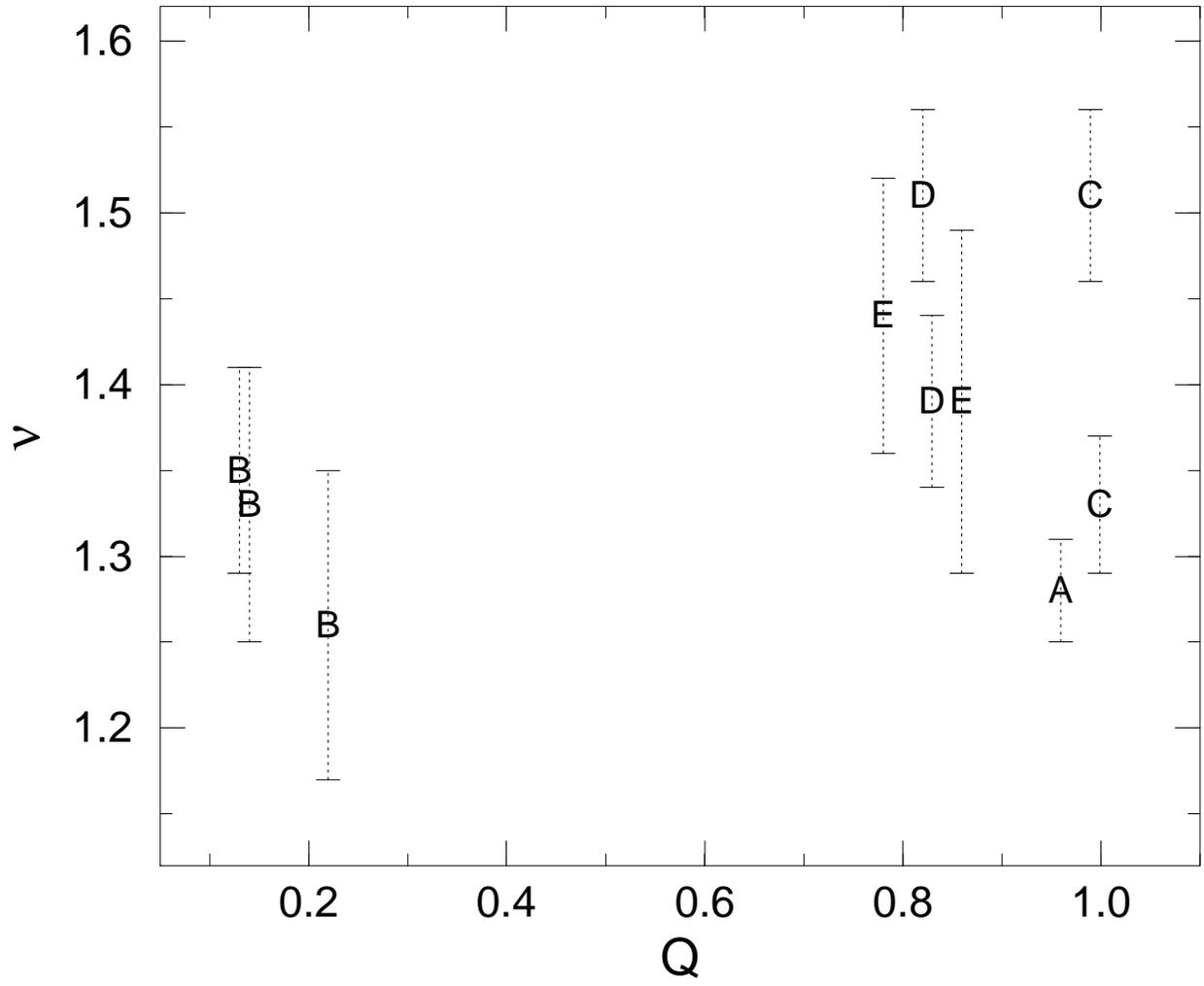}}
  \vspace{1cm}
  \caption{\label{fig:FitResults}
    Results for $\nu$ and their 95\% confidence intervals for the fits
    of the $\alpha_N$ data as reported in Table \ref{tab:fitdata_ELS}.}
\end{figure}

In the ELS data we do not find a clear shift of the crossing point up
to the accuracy of our data as shown in Fig.\ \ref{fig:alpha_W}.
Therefore, an irrelevant scaling variable is probably not necessary
when using the FSS of Sec.\ \ref{sec-Exponent} and we usually set
$n_{\rm i}=0$.  However, the nonlinearities in the $W$ dependence of
$\alpha_N$ require fit functions with $W^3$ terms, i.e., $n_{\rm r}\ge
3$ or $m_{\rm r}\ge 3$. {E.g.}, our choice $n_{\rm r}=3$, $m_{\rm
  r}=1$ yields the fit function
\begin{equation}
  \label{eq-FitFunction}
  \alpha_N=a_{00} + a_{01} N^{1/\nu} w + a_{02} N^{2/\nu} w^2 +
  a_{03} N^{3/\nu} w^3 \quad .
\end{equation}
To achieve a good quality of fit while avoiding higher orders of the
expansions than 3, we use the reduced $W$ interval $[7,11]$ in the
fits. Additionally, the interval $W\in [8,9.25]$ where we have data
for $N=50$ is employed and we vary the $N$ interval in order to find a
possible trend when using larger system sizes only. In Table
\ref{tab:fitdata_ELS} the parameters of the fit function, $\chi^2$,
$Q$ and the results for $W_c$ and $\nu$ with their confidence
intervals are summarized.  We report the best fits obtained for five
combinations of $N$ and $W$ intervals as denoted by the characters A
to E. For completeness, a few worse fits, denoted by small characters,
are added.  First one notices that the quality of fit is rather good
for most cases, i.e., $Q>0.8$. Only for group B, we find smaller
values $Q=0.13 - 0.22$ which nevertheless still indicate a useful fit.
Thus the fit functions describe the data very well.  This can be also
seen in Fig.\ \ref{fig:alpha_W}, where we added the fitted model
functions of fit A. As shown in Fig.\ \ref{fig:scalf_alpha}, the data
collapse onto a single scaling function with two branches
corresponding to the localized and extended regime. This indicates,
together with the divergent scaling parameter shown in the inset of
Fig.\ \ref{fig:scalf_alpha}, clearly the MIT.  The values of the
critical disorder obtained from the different fits are scattered from
$8.54$ to $8.62$ although the 95\% confidence intervals as given by
the LMM are $\pm 0.02$ only.  Apparently, these error estimates do not
characterize the real situation. Thus, we conclude $W_c=8.58\pm 0.06$.
For the critical exponent the situation is even worse as visualized in
Fig.\ \ref{fig:FitResults}. The results scatter from $1.26$ up to
$1.51$.  Some of the 95\% confidence intervals which range from $\pm
0.04$ to $\pm 0.10$ overlap but others are far apart.  Consider for
instance group C: while both fits have a $Q$ value of nearly one,
their estimates for $\nu$ differ by twice the width of the confidence
intervals. This is unexpected since the standard deviation of the
$\alpha_N$ values is taken into account by the LMM. We have also
tested the use of an energy interval containing only 20\% of the
eigenvalues instead of 50\% for the determination of the $\alpha_N$
data, because one might argue that the ELS and thus the critical
behavior depends on $E$ more strongly in the anisotropic case than in
the isotropic system where no changes were found for a large $E$
interval.\cite{BulSK87} However, we find no significant changes of the
results in the anisotropic case, either.

An interesting trend can be seen by considering the mean value of the
fitted $\nu$ within the groups. For A and B it is 1.28 and 1.32, for
the three other groups C,D,E, where the data for smaller system sizes
($N=13,17,21$) are neglected, the mean $\nu$ is 1.42, 1.45, and 1.42.
The fitted critical exponent apparently increases if only larger
system sizes are considered.  This is a hint that in the thermodynamic
limit $\nu$ is probably larger than 1.4. And it also might indicate
that there are finite-size corrections in the $\alpha_N$ data that are
not described by the fit functions. Considering this trend and the
scattering of the results from all fits we conclude from our ELS data
the critical exponent $\nu=1.45\pm0.2$. This is consistent with other
ELS results for the orthogonal case, $\nu=1.34\pm0.10$\cite{HofS94b}
and $1.4\pm0.15$\cite{ZhaK97}. As our results are obtained from more
accurate data due to more samples and larger system sizes, and in the
light of the above discussion, we believe that the error estimates of
Refs.\ \onlinecite{HofS94b,ZhaK97} are too small. Our $\nu$ is small
compared to results from highly accurate TMM studies.  For the present
case of weakly coupled planes $\nu=1.62\pm0.08$ was
obtained.\cite{MilRS99b} For the isotropic case without magnetic field
$\nu=1.54\pm0.08$\cite{Kin94} and $\nu=1.58\pm0.06$\cite{SleO99a} was
determined. However, considering the error bars obtained from the
scattering of the results from different fits, the results are
consistent.

\begin{figure}[p]
  \centerline{\psfig{figure=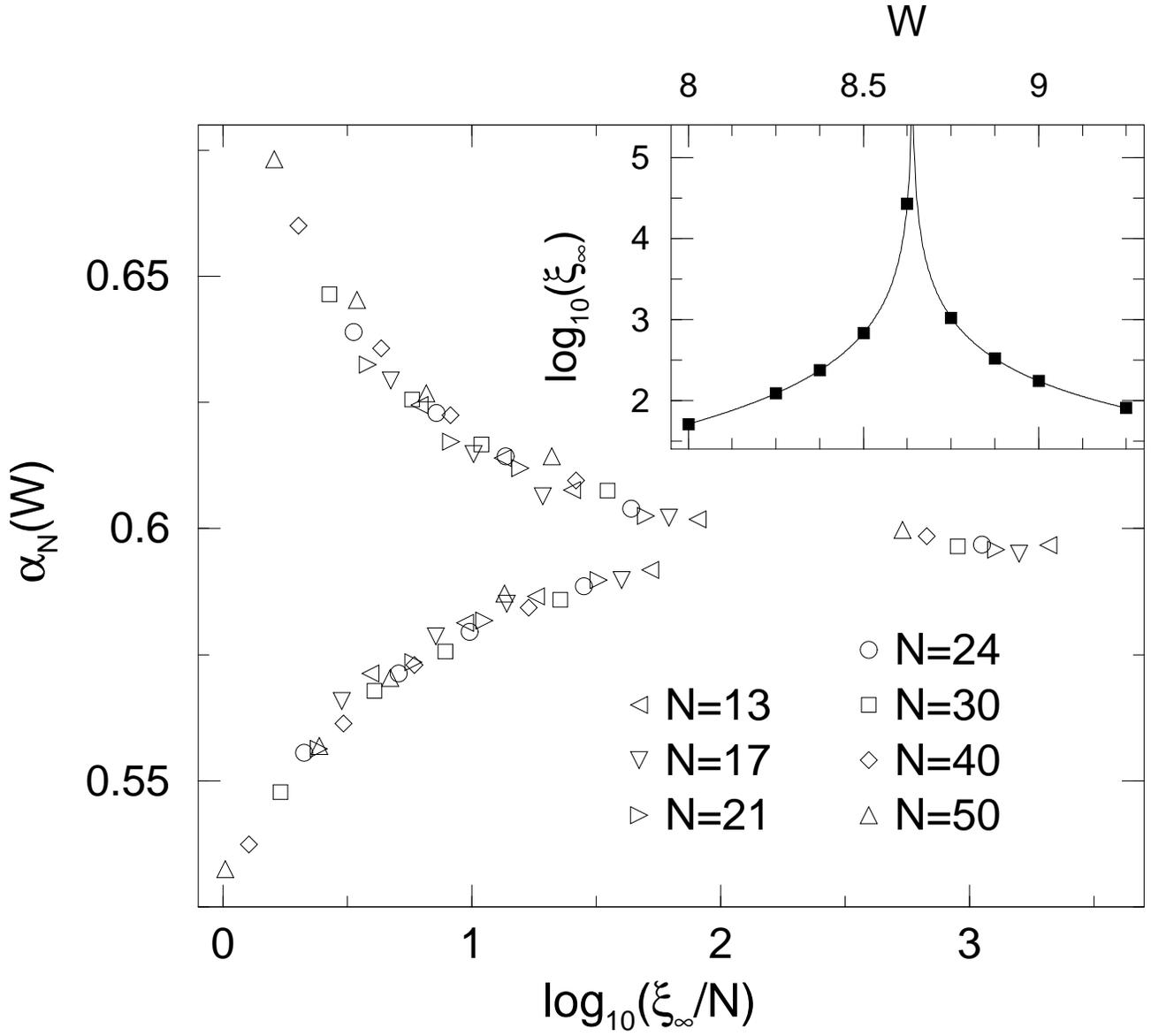}}
  \vspace{1cm}
  \caption{\label{fig:ELS_TMM_scaled}
    $\alpha_N(W)$ for coupled planes with $\gamma=0.9$ scaled with
    $\xi_\infty(W)$, as shown in the inset, taken from a fit of highly
    accurate TMM data with $\nu=1.59$ of
    Ref.~\protect\onlinecite{MilRS99b}.}
\end{figure}

As a further test to check whether the results from TMM and ELS are
compatible, we scaled the $\alpha_N(W)$ data with the scaling
parameter $\xi_\infty(W)$ available from FSS of TMM data with 0.07\%
error.\cite{MilRS99b} As can be seen in Fig.\ 
\ref{fig:ELS_TMM_scaled}, the $\alpha_N$ data collapse reasonably well
onto a single scaling function. Apparently, the $N=50$ data lie
systematically slightly above the scaling function. We remark that in
our calculations, the seed that initializes the random number
generator depends only on the number of the sample but not on the
value of $W$.  The potential energies for the $k$th sample are
obtained by scaling the $k$th sequence of random numbers with $W$.
For a small number of samples, i.e., 10 for $N=50$, this might lead to
the observed systematic shift. Another possible reason is, that the
influence of an irrelevant scaling variable starts to become visible.
But using such a variable in the fits gives no improvement of the
results although the number of parameters increases remarkably. In
group g of Table \ref{tab:fitdata_ELS} we report a few of such fits.
Compared to group B the $Q$ values are much larger. But for the first
two fits the confidence intervals are extremely large and the
estimates for $\nu$ and $W_c$ are nearly meaningless. The third fit
gives essentially the same results as the corresponding fit of group
B. We remark that one has to be very careful when using such
non-linear fits. Even an excellent $Q$ value does not guarantee useful
results for the fitted model parameters.

For completeness we also used the fit method\cite{Hof98} described at
the end of Sec.~\ref{sec-Exponent}. The third-order polynomials are
shown in the inset of Fig.\ \ref{fig:alpha_W} and the scaled data in
Fig.\ \ref{fig:scalf_alpha}. The scaling is almost perfect, which is
not surprising because of the effective smoothing of the data by means
of the cubic fits. The scaling function is very similar to that from
fit f which gives the same critical exponent $\nu\approx 1.3$. But as
mentioned before, it is very difficult to derive a reliable $\nu$ and
its error estimate with this method.

%
%

\section{Summary}
\label{sec-SUM}

In the present work we have studied the metal-insulator transition in
the 3D Anderson model of localization with anisotropic hopping. We
used ELS together with FSS analysis to characterize the MIT.  Our
results indicate in agreement with previous studies that a transition
exists for any anisotropy $\gamma < 1$ for coupled planes and for
coupled chains. We find a power-law decay of the critical disorder
with increasing anisotropy. For coupled chains the decay is found to
be faster than predicted by previous CPA results.\cite{ZamLES96a} A
large part of the present work is devoted to the determination of the
critical exponent $\nu$.  We calculated the integral $\alpha_N(W)$
over the $\Delta_3$ statistics with very large system sizes and high
accuracy for the case of coupled planes with $\gamma=0.9$. In order to
determine $\nu$ we used a method to fit the data introduced by Slevin
and Ohtsuki\cite{SleO99a} which allows for corrections to scaling due
to an irrelevant scaling variable and nonlinearities in the disorder
dependence of the scaling variables. It turns out that it is not
necessary to take into account an irrelevant scaling variable.  By
varying the ranges of system size and disorder we find a large number
of fits, which all describe the data remarkably good.  One expects
that the estimates for $\nu$ from all these fits are consistent within
the error bars.  Unfortunately, the $95\%$ confidence intervals do not
always overlap.  Apparently, the error estimate from the fit procedure
does not reflect the possible fluctuations in $\nu$.  Furthermore, we
find a systematic shift towards larger values of $\nu$ if only the
largest system sizes are considered.  This might indicate the
existence of finite-size corrections not included in the ansatz.
Taken all these into account we conclude $\nu=1.45\pm0.2$. Considering
the accuracy of our data and the large system sizes used, this error
estimate appears surprisingly large when compared to previous
estimates. We presume that most specified errors in similar studies
are somewhat optimistic. Within the error bars, our result is
consistent with highly accurate TMM studies for the GOE
case.\cite{MilRS99b,SleO99a,Kin94} This supports the concept of
universality classes, from which one expects no change of the critical
exponent for anisotropic hopping.

The ELS data at the MIT are independent of the system size, but they
depend on anisotropy. For increasing anisotropy we find a trend from
the isotropic form towards Poisson statistics.  Since it was found
that the critical statistics also depends on boundary conditions and
on the shape of the sample,\cite{PotS98,BraMP98,SchP98} as well as on
the dimensionality,\cite{ZhaK98} one might doubt the validity of the
very attractive concept of a super universality class\cite{Hof98}
characterizing the critical point in all universality classes. That
concept is based on the observation of very similar critical exponents
and agreement of the large-$s$ behavior of $P(s)$ at $W_c$ in the
isotropic Anderson model. We believe that at present the accuracy in
the determination of the critical properties does not allow for an
entirely convincing validation of this concept.

\acknowledgments This work was supported by the DFG within the
Sonderforschungsbereich 393.


%
%

\newpage

\begin{table}[h]
 \begin{center}
  \leavevmode
  \begin{tabular}[h]{|c|c|c|ccc|c|c|c|c|}
    &$N$&$W$&$n_{\rm r}$&$m_{\rm r}$&$n_{\rm i}$&$\chi^2$&$Q$&$W_c$&$\nu$\\
    \hline
    A&$13\cdots 40$ & $7\cdots 11$  & 3 &1 &0&52.6 &0.96 &$8.62\pm0.02$ &$1.28\pm 0.03$\\
    a&$13\cdots 40$ & $7\cdots 11$  & 1 &3 &0&96.6 &0.03 &$8.60\pm0.02$ &$1.41\pm 0.03$\\

    b&$13\cdots 50$ & $8\cdots 9.25$& 3 &1 &0&71.4 &0.095 &$8.59\pm0.02$ &$1.33\pm 0.08$\\
    b&$13\cdots 50$ & $8\cdots 9.25$& 1 &3 &0&78.8 &0.03 &$8.58\pm0.02$ &$1.38\pm 0.06$\\
    B&$13\cdots 50$ & $8\cdots 9.25$& 3 &2 &0&67.7 &0.14 &$8.59\pm0.02$ &$1.33\pm 0.08$\\
    B&$13\cdots 50$ & $8\cdots 9.25$& 2 &3 &0&68.2 &0.13 &$8.60\pm0.02$ &$1.35\pm 0.06$\\
    B&$13\cdots 50$ & $8\cdots 9.25$& 3 &3 &0&62.8 &0.22 &$8.59\pm0.02$ &$1.26\pm 0.09$\\

    C&$24\cdots 40$ & $7\cdots 11$  & 3 &1 &0&11.1 &0.999&$8.60\pm0.02$ &$1.33\pm 0.04$\\
    C&$24\cdots 40$ & $7\cdots 11$  & 1 &3 &0&17.1 &0.99 &$8.58\pm0.03$ &$1.51\pm 0.05$\\

    D&$24\cdots 50$ & $7\cdots 11$  & 3 &1 &0&33.4 &0.83 &$8.55\pm0.02$ &$1.39\pm 0.05$\\
    D&$24\cdots 50$ & $7\cdots 11$  & 1 &3 &0&33.7 &0.82 &$8.54\pm0.02$ &$1.51\pm 0.05$\\

    E&$24\cdots 50$ & $8\cdots 9.25$& 3 &1 &0&21.8 &0.86 &$8.55\pm0.02$ &$1.39\pm 0.10$\\
    E&$24\cdots 50$ & $8\cdots 9.25$& 1 &3 &0&23.7 &0.78 &$8.55\pm0.02$ &$1.44\pm 0.08$\\

    f&$13\cdots 50$ & $7\cdots 11$  & 3 &3 &0&97.8 &0.075 &$8.58\pm0.02$ &$1.33\pm 0.04$\\

    g&$13\cdots 50$ & $8\cdots 9.25$& 1 &3 &1&44.1 &0.83 &$8.31\pm1.31$ &$2.08\pm 6.09$\\
    g&$13\cdots 50$ & $8\cdots 9.25$& 3 &1 &1&37.5 &0.94 &$8.49\pm0.22$ &$0.66\pm 1.08$\\
    g&$13\cdots 50$ & $8\cdots 9.25$& 2 &3 &1&38.1 &0.92 &$8.55\pm0.04$ &$1.35\pm 0.06$\\
  \end{tabular}
  \caption{
    Fit parameters and estimates for $W_c$ and $\nu$ with 95\%
    confidence intervals for the fit of $\alpha_N(W)$ for coupled
    planes with $\gamma=0.9$, $m_{\rm i} = 0$.}
  \label{tab:fitdata_ELS}
\end{center}
\end{table}

%
%

\newpage

\newpage

\newpage

\newpage

\newpage

\newpage

\newpage

\newpage

\end{document}